\documentclass{elsart}
\usepackage{amssymb}
\usepackage{amsbsy}
\journal{Physica A}
\begin{document}
\begin{frontmatter}
\title{Field-Theoretical Analysis of %
Singularities at Critical End Points\thanksref{talkDS}}
\thanks[talkDS]{Talk presented at the Symposium %
on Statistical Physics StatPhys-%
Taipei 99 (Taipei and Hualien, August 1999).}
\author{H. W. Diehl
} \and
\author{M.\ Smock
}
\address{Fachbereich Physik, Universit{\"a}t-Gesamthochschule %
Essen, D-45117 Essen, Federal Republic of Germany}
\begin{abstract}
Continuum models with critical end points are considered
whose Hamiltonian ${\mathcal{H}}[\phi,\psi]$ depends on
two densities $\phi$ and $\psi$.
Field-theoretic methods are used to show the equivalence of
the critical behavior on the
critical line and at the critical end point and to give
a systematic derivation of
critical-end-point singularities like the thermal
singularity $\sim|{t}|^{2-\alpha}$ of the spectator-phase boundary
and the coexistence singularities $\sim |{t}|^{1-\alpha}$ or
$\sim|{t}|^{\beta}$
of the secondary density $\langle\psi\rangle$.
The appearance of a discontinuity eigenexponent
associated with the critical end point is confirmed, and
the mechanism by which it arises in field theory is clarified.
\end{abstract}
\end{frontmatter}
\section{Introduction}
Critical end points are ubiquitous in nature. They occur when a
line of critical temperatures $T_{\mathrm{c}}(g)$, depending on
a nonordering field $g$ such as chemical potential
or pressure, terminates at a line $g_\sigma(T)$
of discontinuous phase transitions
\cite{Gri73,Fis89,FB91}. Two familiar examples are: (i) the
critical end point (CEP)
of a binary fluid mixture where the critical line of demixing
ends on the liquid-gas coexistence curve; (ii) the
CEP of \nuc{4}{He}, the terminus of the lambda line on
the gas-phase boundary. On the critical (or lambda) line
the disordered and ordered phases separated by it become
\emph{identical} critical phases; in the case of a binary fluid
with components A and B,
the disordered phase corresponds to
a homogeneously mixed fluid $\alpha\beta$, and the ordered
ones to an A-rich phase
$\alpha$ and a B-rich phase $\beta$; in the case of \nuc{4}{He},
the disordered and ordered phases
are normalfluid and superfluid, respectively.
A crucial feature of a CEP is that a \emph{critical} phase
\emph{coexists} with a \emph{noncritical}
(`spectator') phase $\gamma$ there.

Although CEPs were encountered
in numerous studies of
bulk and interfacial critical phenomena
\cite[and their references]{Gri73,Fis89,FB91,FU90aFU90b,%
BW76,KGYF81,Wid77Wid85,BF91b,Bar91,RCC92,GD98}
in the past decades,
they have rarely been investigated for their own sake.
This may be due to the expectation
that the critical phenomena at a CEP
should not differ in any significant way
from critical phenomena along
the critical line $T_{\mathrm{c}}(g)$ \cite{Gri73}.
However, recently
it has been pointed out \cite{Fis89,FB91,FU90aFU90b}
that even
the \emph{bulk thermodynamics} of a CEP
should exhibit new critical singularities,
not observable on the critical line. On the basis
of the phenomenological
theory of scaling it was predicted \cite{Fis89}
that the first-order phase boundary $g_\sigma(T)$
should vary
near the CEP $(T=T_{\mathrm{e}},
g=g_\mathrm{e}\equiv g_\sigma(T_{\mathrm{e}}))$ as
\begin{equation}\label{singpb}
g_\sigma(T)\approx
g^{\mathrm{reg}}_\sigma(T)-\frac{X^0_\pm}{(2-\alpha)(1-\alpha)}
\,\left|{t}\right|^{2-\alpha}
\end{equation}
in the limit $t\equiv (T-T_{\mathrm{e}})/T_{\mathrm{e}}\to 0\pm$,
where $g^{\mathrm{reg}}_\sigma(T)$ is \emph{regular} in $T$.
Furthermore, the amplitude ratio $X^0_+/X^0_-$ should be equal
to the usual \emph{universal} (and hence $g$
\emph{independent}) ratio $A_+/A_-$ of specific heat amplitudes
$A_\pm$. These are
defined by writing the specific heat
singularity at constant $g\ne g_e$
on the critical line as $A_\pm(g)\,
\left|T-T_{\mathrm{c}}(g)\right|^{-\alpha}$.
In other words, the singularities displayed by $g_\sigma(T)$ should
be of the same form as those of
the bulk free energy of the disordered ($\alpha\beta$) and
ordered ($\alpha+\beta$) phases near $T_{\mathrm{c}}(g)$.

The phenomenological scaling arguments leading to (\ref{singpb}) can
be extended in a straightforward fashion to determine the
singularities $\rho_g$, the thermodynamic
density conjugate to the nonordering field $g$, should display
as the CEP is approached along the coexistence
boundary \cite{Wil97aWil97b}.
They yield the singular part
\begin{equation}\label{singdens}
 \rho^{\mathrm{sing}}_g\equiv \rho_g-
\rho^{\mathrm{reg}}_g(T)\approx 
U^0_\pm\,\left|{t}\right|^{\beta}
+V^0_\pm\,\left|{t}\right|^{1-\alpha}\;.
\end{equation}
Having in mind binary fluid mixtures, we take the CEP to lie on
the liquid (rather than the gas) side of the coexistence boundary.
The quantity $\rho_g$ may be identified
as the total density of the fluid.
 For a hypothetical \emph{symmetric} binary fluid
whose properties are invariant with regard to simultaneous
interchange of its two constituents A and B
and their respective chemical potentials $\mu_{\mathrm{A}}$
and $\mu_{\mathrm{B}}$, the amplitudes $U^0_\pm$ would vanish.
More generally, this would be true
for systems that are describable by a continuum Hamiltonian
which is \emph{even} in the order parameter field $\phi$. Just as
$X^0_+/X^0_-$, the ratios $V^0_+/V^0_-$ as well as $U^0_+/U^0_-$
(if $U^0_\pm\neq 0$) are \emph{universal} and can be expressed in terms of
standard universal amplitude combinations \cite{Wil97aWil97b}.

The  $|{t}|^{2-\alpha}$ singularity of (\ref{singpb}) has been
checked by Monte Carlo
calculations \cite{Wil97aWil97b} and verified
for exactly solvable spherical models \cite{BF91b,Bar91};
the $|{t}|^{1-\alpha}$
singularity of (\ref{singdens})
is consistent with the jump in the slope of $\rho_g(T)$
found in mean field and density functional calculations
\cite{RCC92,GD98} and
has also been seen in Monte Carlo simulations
\cite{Wil97aWil97b,WSN98}. 

Here we will address the issue of CEP singularities
via the \emph{field-theoretic renormalization group (RG) method}.
This approach is known to provide both
a conceptually reliable basis of the modern theory of
critical phenomena as well as powerful calculational
tools (see, e.g., \cite{DG76,ZJ96}). Surprisingly, it has not yet
been applied with much success to the study of
CEPs. We are aware of only one such work
that goes beyond the
Landau approximation, an
$(\epsilon=4-d)$-expansion study of
a scalar  $\phi^8$ model with negative $\phi^6$ term  \cite{ZAGJ82}.
Its one-loop result is
that the critical line and the CEP are controlled by the same,
standard $O(\epsilon)$ fixed point. Unfortunately,
the model investigated has \emph{rather special features}:
its first-order line does {\em not\/}
extend into the disordered phase;
as its CEP is approached from the disordered phase, the
order parameter $\langle\phi\rangle$ becomes critical
\emph{and} exhibits a
jump to a nonvanishing value upon entering the ordered phase; and
no critical fluctuations occur in its ordered phase.
Hence it clearly \emph{does not reflect} the typical CEP situation
in which the two-phase coexistence surface
bounded by the critical line $T_{\mathrm{c}}(g)$ meets the
spectator phase boundary in a triple line;
its applicability appears to be quite limited.

One should also note that the above RG scenario
\emph{differs} from the one found in
position-space RG calculations \cite{BW76,KGYF81}
of lattice models with conventional CEPs.
In the latter scenario the critical line and the CEP
are mapped onto \emph{separate} fixed points, where the CEP fixed point
has two relevant RG eigenexponents that are identical to those of
the former, plus the additional one $y=d$, characteristic of
discontinuity fixed points \cite{NN75}, but absent in \cite{ZAGJ82}.

We conclude that systematic field-theoretic
RG studies of appropriate models are urgently
needed. A first obvious goal one would hope to
achieve is a systematic derivation of the
singularities in (\ref{singpb}) and (\ref{singdens}). 
This involves showing the equivalence of critical behavior
at the CEP and on the critical line.%
\footnote{In their excellent  survey \cite{FB91} of the present state
of the theory of CEP singularities, Fisher and Barbosa
warn that
this equivalence,
with matching critical spectra of the %
corresponding two fixed points, need not
be an invariable rule, even though their work confirms it, just as
our own.}
Provided the above RG scenario with two separate fixed points prevails,
one must prove that the associated critical spectra match, demonstrate
the existence of the discontinuity eigenexponent $y=d$ and
clarify its significance.

We have recently carried out such an investigation.
In the sequel, we will briefly describe the main steps
of our procedure and our findings. A more detailed exposition of our
work will be given elsewhere \cite{SD99}.

\section{Models}

First, we must choose an appropriate continuum model.
Natural candidates are models whose Hamiltonian
$\mathcal{H}[\phi,\psi]$ depends on \emph{two} fluctuating
densities: a (primary) order parameter field
$\phi(\boldsymbol{x})$ and
a secondary (noncritical) density $\psi(\boldsymbol{x})$.
The form of $\mathcal{H}$ can be
guessed on purely phenomenological grounds,
but can also be derived by starting from an appropriate
lattice model, such as the Blume-Emery-Griffiths model \cite{BEG71}
on a $d$-dimensional simple cubic lattice.
This is a classical spin $S=1$ model with Hamiltonian
\begin{eqnarray}\label{HBEG}
{\mathcal H}_\mathrm{BEG}[\boldsymbol{S}]&=&-\sum_{\langle \boldsymbol{i},\boldsymbol{j}\rangle}
\left[J\,S_{\boldsymbol{i}}\,S_{\boldsymbol{j}}
+K\,S_{\boldsymbol{i}}^2\,S_{\boldsymbol{j}}^2
+L\left(
S_{\boldsymbol{i}}^2\,S_{\boldsymbol{j}}
+S_{\boldsymbol{i}}\,S_{\boldsymbol{j}}^2\right)\right]
\nonumber\\
&&\mbox{}-\sum_{\boldsymbol{i}}\left(
H\,S_{\boldsymbol{i}}+D\,S_{\boldsymbol{i}}^2 \right),
\qquad\qquad S_{\boldsymbol{i}}=0,\pm 1\,,
\end{eqnarray}
where  $\langle \boldsymbol{i},\boldsymbol{j}\rangle$
indicates summation over nearest-neighbor pairs of sites.
We presume the interaction constants $K$ and $J$ to be positive (`ferromagnetic'), and $L\ge 0$. The quantities
$H$ and $D$ correspond respectively to even and odd linear
combinations of the chemical potentials
$\mu_\mathrm{A}$ and $\mu_\mathrm{B}$ \cite{SL75a}.

Performing a Gaussian (`Kac-Hubbard-Stratonovich')
transformation with respect
to both $\{S_{\boldsymbol{i}}\}$ and $\{S^2_{\boldsymbol{i}}\}$,
one can map the model (\ref{HBEG}) exactly on a lattice field theory
with fields $\phi_{\boldsymbol{i}}\in\Rset$
and $\psi_{\boldsymbol{i}}\in\Rset$ \cite{SD99}.
To make a continuum approximation, we replace these
by smoothly interpolating
fields $\phi({\boldsymbol{x}})$ and $\psi({\boldsymbol{x}})$, and
Taylor expand nearby differences
$\phi_{\boldsymbol{i}}-\phi_{\boldsymbol{j}}$ about their midpoint
$(\boldsymbol{i}+\boldsymbol{j})/2$. We thus arrive at a continuum
model with the Hamiltonian
\begin{equation}\label{ham}
{\mathcal{H}}[\phi,\psi]={\mathcal{H}}_1[\phi]+{\mathcal{H}}_2[\psi]
+{\mathcal{H}}_{12}[\phi,\psi]\;,
\end{equation}
\begin{equation}\label{H1}
{\mathcal{H}}_1[\phi]=\int\!d^dx\left[
\frac{A}{2}\left(\nabla\phi\right)^2+\frac{a_2}{2}\, \phi^2+
\frac{a_4}{4}\,\phi^4-h\phi
\right],
\end{equation}
\begin{equation}\label{H2}
{\mathcal{H}}_2[\psi]=\int\!d^dx\left[
\frac{B}{2}\left(\nabla\psi\right)^2+\frac{b_2}{2}\, \psi^2+
\frac{b_4}{4}\,\psi^4-g\psi
\right],
\end{equation}
\begin{equation}\label{H12}
{\mathcal{H}}_{12}[\phi,\psi]=\int\!d^dx 
\left[\psi\left(d_{11}\,\phi+
\frac{d_{21}}{2}\,\phi^2\right)
+\triangle\psi \left(
e_{11}\,\phi+\frac{e_{21}}{2}\,\phi^2\right)
\right].
\end{equation}
The $\phi^3$ and $\psi^3$ terms have been eliminated by
shifts $\phi(\boldsymbol{x})
\to\phi(\boldsymbol{x}) +\phi_0$ and
$\psi(\boldsymbol{x})
\to\psi(\boldsymbol{x}) +\psi_0$. Monomials of higher order
and higher-order gradient terms have been dropped.

The terms retained in $\mathcal{H}$ require explanation.
Consider, first, the case of a \emph{symmetric} CEP, in which $\mathcal{H}[-\phi,\psi]=\mathcal{H}[\phi,\psi]$,
i.e., $h=d_{11}=e_{11}=0$.
Owing to this symmetry, $\phi$ and $\psi$ do not `mix' and hence may
be chosen as the fields that become critical or remain noncritical
at the CEP, respectively. To assess the relevance of contributions to
$\mathcal{H}$ via power counting, the coefficients
$A$ and $b_2$ should be taken dimensionless.
Thus $\phi$ has the usual momentum dimension
$[\phi]=(d-2)/2$, while $[\psi]=d/2$. In
$\mathcal{H}_1[\phi]$, we have kept all monomials of the standard
$\phi^4$ Hamiltonian, namely those (except $\phi^3$)
having coefficients with nonnegative momentum dimensions for
$\epsilon \ge 0$. For the remaining interaction constants, one
finds $[g]=-[e_{21}]=2-\epsilon/2$, $[d_{21}]={\epsilon/ 2}$,
$[B]=-2$, and $[b_4]=\epsilon-4$. This suggests that
$B$, $e_{21}$, and $b_4$ may be expected to be irrelevant in the RG sense
and hence can be set to zero. If we did this, $\mathcal{H}$ would
reduce to the Hamiltonian of
the dynamic model C \cite{HH77}; it would be quadratic in $\psi$,
so $\psi$ could be integrated out exactly. The resulting
effective Hamiltonian would be
identical to $\mathcal{H}_1[\phi]$, up to a change of its parameters
$a_2$ and $a_4$, and an overall constant.

The terms $\propto B$, $e_{11}$, and  $e_{21}$
have been introduced because
they play a role in the analysis of inhomogeneous states with
a liquid-gas interface%
\footnote{For $B>0$, %
the mean-field correlation length of $\psi$ and hence the width %
of the interface region of classical kink solutions for $\psi$ are nonzero.
The terms $\propto e_{11}$ and $e_{21}$ are significant 
 for relating the problem of critical
adsorption of the $\alpha\beta$-phase at the
$\alpha\beta|\gamma$-interface to a wall problem.}
\cite{SD99}.
Since our main focus here is on bulk critical behavior, we
can indeed set $B=e_{21}=0$ in the sequel. However, $b_4$ must
\emph{not} be set to zero because, then,
we would \emph{not} be able to describe $\alpha\beta$-$\gamma$ coexistence, nor
would the model have a CEP.

\section{Landau theory and beyond}

Application of the Landau approximation to the model
(\ref{ham})--(\ref{H12}) yields a phase
diagram with a CEP and the correct topology,
provided its parameter values are in the appropriate range 
\cite{RCC92,WSN98,SD99}. With the choices $a_2<0$, $d_{21}>0$
(aside from $a_4>0$ and $b_4>0$), one finds a critical line with
\begin{equation}
\psi=\psi_\mathrm{c}\equiv -{a_2/d_{21}}\;,\quad
b_2>b_{2\mathrm{e}}\equiv\frac{{d_{21}^2}}{2\,a_4} - \frac{b_4\,a_2^2}{d_{21}^2}\;,
\end{equation}
and $g=g_\mathrm{c}=b_2\psi_\mathrm{c}+b_4\psi_\mathrm{c}^3$
that is truncated  by the
liquid-gas coexistence boundary
at the CEP $b_2=b_{2\mathrm{e}}$, $\psi=\psi_\mathrm{c}$
(cf.\ case (a) in Figs.~8 and 9 of \cite{RCC92}).
A detailed exposition of the Landau theory, with results for the
phase boundaries and equilibrium values of
$\phi$ and $\psi$ in the various phases will be given in \cite{SD99}.

To go beyond Landau theory, we use perturbation theory in combination
with the RG.%
\footnote{Implicit in our analysis is the well-founded assumption that
the CEP, the critical line, and the first-order line $g_\sigma$ will survive
the inclusion of fluctuation corrections. }
Writing 
$\psi=\psi_\mathrm{ref}+\check\psi$, we
expand $\mathcal{H}$ about a reference
value $\psi_\mathrm{ref}$,
which we take as the mean-field value of $\psi$ at a reference point
in the $\alpha\beta$ phase away from the critical line. This gives
\begin{equation}\label{Hdecomp}
\mathcal{H}[\phi,\psi]=
\mathcal{H}[0,\psi_\mathrm{ref}]
+\mathcal{H}'[\phi,\check\psi;\psi_\mathrm{ref}]\;,
\end{equation}
\begin{equation}\label{Hprime}
\mathcal{H}'[\phi,\check\psi]=
\int\!d^dx\left[\frac{A}{2}\left(\nabla\phi\right)^2 +\sum_{k=2,4}
\frac{\check a_k}{k}\,{\phi}^k+\sum_{l=1}^4\frac{\check b_l}{l}
\,{\check\psi}^l
+\frac{1}{2}\,d_{21}{\phi}^2\check\psi\right]\,,
\end{equation}
with $\check a_2=a_2+d_{21}\,\psi_\mathrm{ref}$, $\check a_4=a_4$,
$\check b_1=-g+b_2\,\psi_\mathrm{ref}+b_4(\psi_\mathrm{ref})^3$,
$\check b_2=b_2+3b_4(\psi_\mathrm{ref})^2$,
$\check b_3=3b_4\,\psi_\mathrm{ref}$, and $\check b_4=b_4$,
where $\check b_2>0$.

If $\mathcal{H}'$ is taken into account by perturbation theory,
the critical and first-order lines, and hence the CEP, get shifted.
When studying the behavior near these lines,
one must use their corresponding new locations that are compatible
with the level of approximation. Suppose  that a point $(T_{\mathrm{c}}(g_{\mathrm{c}}),g_{\mathrm{c}})$
on the critical line is approached, which may be the CEP ($g_{\mathrm{c}}=g_{\mathrm{e}}$).
Let us ignore the $\check\psi^3$ and $\check\psi^4$ terms for the present.
Then $\mathcal{H}$ becomes quadratic in $\check\psi$, so that $\check\psi$
can be integrated out exactly. The resulting effective Hamiltonian
$\mathcal{H}_\mathrm{eff}[\phi]$ is given by
$\mathcal{H}_1[\phi]$, with the replacements 
$a_2\to \check a_2 -(\check b_1\,d_{21}/\check b_2)$ and
$a_4\to \check a_4-d_{21}^2/{2\check b_2}$, plus a $\phi$-independent
term  $\int\!d^dx\,f_\mathrm{G}(\check b_2,\check{b}_1)$ corresponding to
 the free energy of the Hamiltonian
$\mathcal{H}_\mathrm{G}[\check\psi]
=\int\!d^dx\,[(\check b_2/2)\,\check\psi^2+\check b_1\,\check\psi]$.
As usual, we may presume that
parameters such as $\check a_2$, $\check a_4$ have a Taylor expansion in
$t=(T-T_{\mathrm{c}}(g_{\mathrm{c}}))/T_{\mathrm{c}}(g_{\mathrm{c}})$ and $\delta g=(g-g_{\mathrm{c}})/g_{\mathrm{c}}$
near $(T_c,g_c)$. Both $\mathcal{H}[0,\psi_\mathrm{ref}]$
as well as $f_\mathrm{G}$ are regular 
at $(T_c,g_c)$. Hence the singular part of the total free energy results
solely from $\mathcal{H}_\mathrm{eff}$.
That it has the usual scaling form near criticality
can be demonstrated via a standard field-theoretic
RG analysis, either of $\mathcal{H}_\mathrm{eff}$ directly,
or of the equivalent model-C-type Hamiltonian from which it
originated. Thus the critical behavior
on the critical line is the same as at the CEP,
provided our omission of the $\check\psi^3$
and $\check\psi^4$ terms was justified.

It is instructive to consider these nonlinearities first
in the case $d_{21}=0$ of a massive field $\check\psi$
decoupled from $\phi$.
Let $\mathfrak{G}_\mathrm{tra}[\Psi]$
be the generator of transformations of Hamiltonians
$\mathcal{H}[\check\psi]$ induced by
the change of variable $\check\psi\to\check\psi+\Psi$:
\begin{equation}
\mathfrak{G}_\mathrm{tra}[\Psi]\,\mathcal{H}[\check\psi]
\equiv\int\!d^dx
\left[\Psi(\boldsymbol{x})
\frac{\delta\mathcal{H}[\check\psi]}{\delta\check\psi(\boldsymbol{x})}
-\frac{\delta\Psi(\boldsymbol{x})}{\delta\check\psi(\boldsymbol{x})}\right].
\end{equation}
Choosing
\begin{equation}\Psi_k=\frac{\check b_k}{k\,\check b_2}\left[\check\psi^{k-1}+\delta(\boldsymbol{0})\,
\frac{k-1}{\check b_2}\,\check\psi^{k-3}\right],\quad k=3,4,
\end{equation}
gives
\begin{equation}\label{psi34red}
\frac{\check b_k}{k}\int\!d^dx\,\check\psi^k=
\mathfrak{G}_\mathrm{tra}\!\left[\Psi_k\right]\,
\mathcal{H}_\mathrm{G}[\check\psi]
+\delta_{k,4}\,c_4\int\!d^dx\;,\;\;k=3,4\,,
\end{equation}
with $c_4=-3\check b_4\,[\delta(\boldsymbol{0})/2\check b_2]^2$,
where $\delta(\boldsymbol{0})=\int d^dq/(2\pi)^d$ is a cutoff-dependent
constant. Hence, at the \emph{Gaussian} fixed point $\mathcal{H}_\mathrm{G}[\check\psi]$,
the $\check\psi^3$ and $\check\psi^4$ terms correspond
to a \emph{redundant} operator, and a
\emph{redundant} operator plus a constant, respectively.
More generally, Wegner
\cite[Sec.\ III.G.2]{Weg76} has shown that
\emph{any translationally invariant
local operator} can be represented as a redundant operator plus
a constant at such a noncritical Gaussian fixed point.

To generalize these considerations to the case
$\d_{21}\ne 0$ with $\check{b}_1\ne 0$, we insert
\begin{equation}
\mathfrak{G}_\mathrm{tra}\!\left[\Psi_k\right]\,
\mathcal{H}_\mathrm{G}[\check\psi]=
\mathfrak{G}_\mathrm{tra}\!\left[\Psi_k\right]\,
\mathcal{H}'[\phi,\check\psi]-\int\!d^dx\,\left(\frac{d_{21}}{2}\,
\phi^2+\check{b}_1\right)\Psi_k
\end{equation}
into (\ref{psi34red}). The result tells us that,
to first order in $\check b_3$,
the $\check\psi^3$ term is equivalent to
shifts of $\check a_2$, $\check b_2$, and the 
constant part of $\mathcal{H}$, plus a generated
$\phi^2\,\check\psi^2$ contribution. Likewise, the
$\check\psi^4$ term corresponds
(to order $\check b_4$) to the generation
of $\phi^2\,\check\psi^3$ and $\check\psi^3$ contributions,
and shifts of $d_{21}$ and $\check{b}_1$.
Owing to the high naive dimensions $2(d-1)$ and $(5d-4)/2$ of
the operators
$\phi^2\,\check\psi^2$ and $\phi^2\,\check\psi^3$, we may trust that
both produce only irrelevant corrections and hence may be dropped.
Consequently, the effects of the 
$\check\psi^3$ and $\check\psi^4$ terms can be absorbed
through shifts of the parameters $\check a_2,\ldots,\check b_1$
of the Hamiltonian with $\check b_3=\check b_4=0$, apart from
irrelevant corrections. This means a change of
the locations of the critical and first-order lines,
and a corresponding adjustment of, e.g., the temperature scaling field.
In summary, we arrive at an effective $\phi^4$ Hamiltonian $\mathcal{H}_\mathrm{eff}[\phi]$,
\emph{irrespective of whether the critical line or the CEP
is approached}.

Next, we turn to the issue of the \emph{discontinuity eigenexponent} $y=d$.
Consider changes
$g_\mathrm{e}\to g_\mathrm{c}=g_\mathrm{e}+\delta g$,
$\check a_{2\mathrm{e}}\to \check a_{2\mathrm{c}}=
\check a_{2\mathrm{e}}+\delta a_2$, away from the CEP
such that the theory \emph{remains critical}.
As we have seen above, varying $g$ (i.e., $\check b_1$)
alters the configuration-independent part of $\mathcal{H}$, i.e.,
the coefficient, $ \mu_0$, of the `volume operator' $\int d^dx$. Since
$\mu_0$ trivially scales with the exponent $d$ under RG transformations,
it is formally relevant; but it does not contribute to the critical
behavior at continuous phase transitions. Wegner \cite{Weg76} 
therefore proposed to call it `special' scaling field. The above variation
within the critical manifold implies a change
$\delta\mu_0\sim (T_\mathrm{c}- T_\mathrm{e})$ of $\mu_0$.
This is the analog of the eigenperturbation with eigenexponent
$y=d$ found at the CEP fixed point in position-space RG calculations \cite{BW76,KGYF81}.

The above analysis of the symmetric CEP can be extended to
the \emph{nonsymmetric} case, in which the $\phi\to-\phi$
symmetry of $\mathcal{H}[\phi,\psi]$ is broken \cite{SD99}.
As expected, the principal modifications are of a geometrical
nature: the first-order surface bounded by the critical line
is no longer confined to the $h=0$ plane, and the
ordering field $h$ and the temperature variable $t$ mix
in the scaling fields.

Our  analysis confirms that the singular part of the free energy
of the $\alpha\beta$ and $\alpha+\beta$ phases
has the usual scaling form anticipated in the phenomenological
scaling  theory, both on the critical line and at the CEP.
Hence it is clear that the limiting forms (\ref{singpb}) and (\ref{singdens})
must hold and can be derived in a similar fashion as
in phenomenological investigations \cite{Fis89,Wil97aWil97b}.
We will therefore restrict ourselves here to a few remarks.
The singularity (\ref{singpb}) in $g_\sigma(T)$ follows by exploiting the
equality of the free energies (grand potential) of the liquid ($\alpha\beta$, $\alpha$,
$\beta$)  and the spectator ($\gamma$) phases at coexistence.
To derive the behavior of the nonordering density $\rho_g=\langle\psi\rangle$
near the CEP, we have generalized the 
field-theoretic RG analysis
\cite{BdD75} to the $g$-dependent model-C-type Hamiltonian that result
in the symmetric and nonsymmetric cases. In the nonsymmetric case $\psi$
does not only couple to the energy density ($\sim |t|^{1-\alpha}$),
but also to the order parameter, which produces the
additional $|t|^\beta$ singularity
in (\ref{singdens}).

\begin{ack}
We gratefully acknowledge enjoyable discussions with
B.\ N.\ Shalaev, helpful correspondence
with R.\ K.\ P.\ Zia, and
partial support
by the Deutsche Forschungsgemeinschaft (DFG)
via Sonderforschungsbereich 237 and the Leibniz program.
\end{ack}

\end{document}